\documentstyle[12pt]{article}

\newcommand{\Li}[2]{{\mbox{Li}}_{#1}\left(#2\right)}

\newcommand{\be}{\begin{equation}}
\newcommand{\ee}{\end{equation}}
\newcommand{\bea}{\begin{eqnarray}}
\newcommand{\eea}{\end{eqnarray}}
\newcommand{\ep}{\varepsilon}
\newcommand{\Int}{\int\limits}

\begin{document}
 \thispagestyle{empty}
 \begin{flushright}
 {\tt University of Bergen, Department of Physics}    \\[2mm]
 {\tt Scientific/Technical Report No.1994-03}    \\[2mm]
 {\tt ISSN 0803-2696} \\[9mm]
 {February 1994}           \\
\end{flushright}
% \vspace{1cm}
 \vspace*{3cm}
 \begin{center}
 {\bf \large
New results for two-loop off-shell three-point diagrams
}
 \end{center}
 \vspace{1cm}
 \begin{center}
 Natalia I.~Ussyukina\footnote{E-mail address: ussyuk@compnet.msu.su
 }  \\
 \vspace{.3cm}
 {\em
 Institute for Nuclear Physics, Moscow State University, \\
 119899, Moscow, Russia
     }
 \end{center}
 \begin{center}
 and
 \end{center}
 \begin{center}
 {Andrei I.~Davydychev\footnote{On leave from
 Institute for Nuclear Physics,
 Moscow State University, 119899, Moscow, Russia.
 E-mail address: davyd@vsfys1.fi.uib.no}}\\
 \vspace{.3cm}
 {\em Department of Physics, University of Bergen,\\
      All\'{e}gaten 55, N-5007 Bergen, Norway}
 \end{center}
 \hspace{3in}
 \begin{abstract}
A number of exact results for two-loop three-point diagrams with massless
internal particles and arbitrary (off-shell) external momenta
are presented. Divergent contributions are calculated in the framework
of dimensional regularization.
 \end{abstract}

\newpage
 \setcounter{page}{2}

{\bf 1.} There are several reasons why evaluation of loop diagrams
with massless internal particles is important, namely:
(i) some particles are really massless (photons, gluons);
(ii) masses of some particles can be neglected in high-energy
processes;
(iii) such results are needed for constructing the asymptotic expansion
of massive diagrams in the large-external-momenta limit;
(iv) examination of such diagrams helps us to develop and apply new
methods of loop calculations.

There are two basic ``topologies'' of the three-point two-loop
diagrams: planar (fig.~1a) and non-planar (fig.~1b) ones.
It should
be noted that the diagram in fig.~1b is symmetric with respect to
all external lines, while the diagram in fig.~1a is symmetric with respect to
the two lower lines only and therefore should be considered
together with two other diagrams, corresponding to permutations
of external momenta. In realistic calculations, however,
we can also get some numerators (connected with spinor or vector
structure of the particles). As a result, some of the denominators
can cancel in separate terms corresponding to the decomposition in
terms of the scalar integrals (some examples can be found in
ref.~\cite{PV}). Such integrals
(that can be obtained from the diagrams of the types shown in fig.~1(a,b)
when some of denominators are removed) are shown in fig.~1(c-f).
All of them are shown for the case when they are symmetric with respect
to the two lower lines, and the results for ``rotated'' diagrams
can be obtained by permutations of the external momenta (as in the
case of the diagram in fig.~1a). The diagrams in fig.~1(d,e,f) are
divergent, and we shall use dimensional regularization \cite{dimreg}
(with the space-time dimension $n=4-2\ep$). The main difficulty is
that we need to evaluate arising one-loop triangle diagrams up
to terms of order $\ep$.

\newcommand{\parta}
 {\setlength {\unitlength}{0.9mm}
 \begin{picture}(26,54)(0,0)
 \put (13,48) {\line(0,1){6}}
 \put (13,48) {\line(-1,-3){12}}
 \put (13,48) {\line(1,-3){12}}
 \put (8,33)  {\line(1,0){10}}
 \put (3,18)  {\line(1,0){20}}
 \put (13,48) {\circle*{1}}
 \put (8,33)  {\circle*{1}}
 \put (18,33) {\circle*{1}}
 \put (3,18)  {\circle*{1}}
 \put (23,18) {\circle*{1}}
 \put (10,3)  {\makebox(0,0)[bl]{\large (a)}}
 \end{picture}}
\newcommand{\partb}
 {\setlength {\unitlength}{0.9mm}
 \begin{picture}(26,54)(0,0)
 \put (13,48) {\line(0,1){6}}
 \put (13,48) {\line(-1,-3){12}}
 \put (13,48) {\line(1,-3){12}}
 \put (8,33)  {\line(1,-1){15}}
 \put (3,18)  {\line(1,1){15}}
 \put (13,48) {\circle*{1}}
 \put (8,33)  {\circle*{1}}
 \put (18,33) {\circle*{1}}
 \put (3,18)  {\circle*{1}}
 \put (23,18) {\circle*{1}}
 \put (10,3)  {\makebox(0,0)[bl]{\large (b)}}
 \end{picture}}
\newcommand{\partc}
 {\setlength {\unitlength}{0.9mm}
 \begin{picture}(26,54)(0,0)
 \put (13,48) {\line(0,1){6}}
 \put (13,48) {\line(-1,-3){12}}
 \put (13,48) {\line(1,-3){12}}
 \put (13,48) {\line(0,-1){30}}
 \put (3,18)  {\line(1,0){20}}
 \put (13,48) {\circle*{1}}
 \put (13,18)  {\circle*{1}}
 \put (3,18)  {\circle*{1}}
 \put (23,18) {\circle*{1}}
 \put (10,3)  {\makebox(0,0)[bl]{\large (c)}}
 \end{picture}}
\newcommand{\partd}
 {\setlength {\unitlength}{0.9mm}
 \begin{picture}(24,54)(0,0)
 \put (12,48) {\line(0,1){6}}
 \put (12,41) {\circle{14}}
 \put (12,34) {\line(-1,-2){11}}
 \put (12,34) {\line(1,-2){11}}
 \put (3,16)  {\line(1,0){18}}
 \put (12,48) {\circle*{1}}
 \put (12,34) {\circle*{1}}
 \put (3,16)  {\circle*{1}}
 \put (21,16) {\circle*{1}}
 \put (9,3)  {\makebox(0,0)[bl]{\large (d)}}
 \end{picture}}
\newcommand{\parte}
 {\setlength {\unitlength}{0.9mm}
 \begin{picture}(26,54)(0,0)
 \put (13,48) {\line(0,1){6}}
 \put (13,48) {\line(-1,-3){12}}
 \put (13,48) {\line(1,-3){12}}
 \put (4,21)  {\line(1,0){18}}
 \put (17,21) {\oval(10,10)[b]}
 \put (13,48) {\circle*{1}}
 \put (4,21)  {\circle*{1}}
 \put (22,21) {\circle*{1}}
 \put (12,21)  {\circle*{1}}
 \put (10,3)  {\makebox(0,0)[bl]{\large (e)}}
 \end{picture}}
\newcommand{\partf}
 {\setlength {\unitlength}{0.9mm}
 \begin{picture}(26,54)(0,0)
 \put (13,48) {\line(0,1){6}}
 \put (13,48) {\line(-1,-3){12}}
 \put (13,48) {\line(1,-3){12}}
 \put (6,27)  {\line(1,0){14}}
 \put (13,27) {\oval(14,14)[b]}
 \put (13,48) {\circle*{1}}
 \put (6,27)  {\circle*{1}}
 \put (20,27) {\circle*{1}}
 \put (10,3)  {\makebox(0,0)[bl]{\large (f)}}
 \end{picture}}
\begin{figure}[bth]
\[
\begin{array}{cccccc}
\parta & \partb  & \partc & \partd & \parte & \partf
\end{array}
\]
\caption{}
\end{figure}

The remainder of the paper is organized as follows. In section~2 we
collect some useful formulae for one-loop triangles. In section~3
we consider planar diagrams (shown in fig.~1(a,c)), while section~4
deals with the non-planar graph (fig.~1b). Then, section~5
examines the contributions of fig.~1(d,e,f). In section 6 (conclusion)
we discuss the results.

\vspace{3mm}

{\bf 2.} In this section we shall briefly summarize some useful formulae
for one-loop triangle diagrams. We shall employ them below,
when evaluating all two-loop-order contributions.

The scalar one-loop three-point Feynman integral with ingoing momenta
$p_1, p_2, p_3$ (such that $p_1+ p_2+ p_3=0$) is defined,
in $n$ dimensions, as
\be
\label{defJ}
J (n; \; \nu_1  ,\nu_2  ,\nu_3 | \; p_1^2, p_2^2, p_3^2 ) \equiv \int
 \frac{\mbox{d}^n r}{ ((p_2 -r )^2)^{\nu_1}  ((p_1 +r )^2)^{\nu_2}
      (r^2)^{\nu_3} } .
\ee
The powers of propagators, $\nu_i$,
can be non-integer if we employ analytic or dimensional regularization
schemes (in the latter case we can obtain $\nu$'s that depend on $n$
as a result of previous loop integrations). Here and below the ``causal''
prescription $(p^2)^{-\nu} \leftrightarrow (p^2+\mbox{i}0)^{-\nu}$
is understood. Usually we shall omit the arguments $p_1^2, p_2^2, p_3^2$
in $J$.

The Feynman parametric representation of (\ref{defJ}) is
\be
\label{Fp}
J(n; \; \nu_1  ,\nu_2  ,\nu_3  ) = \pi^{n/2} \mbox{i}^{1-n}
\frac{\Gamma \left( \sum \nu_{i} - n/2 \right) }{\prod \Gamma(\nu_{i})}
\Int_0^1 \!\! \Int_0^1 \!\! \Int_0^1
\frac{\prod \alpha_{i}^{\nu_{i} -1} \mbox{d}\alpha_{i} \;\;
            \delta \left( \sum \alpha_{i} -1 \right)}
     {\left( \alpha_1 \alpha_2 p_3^2 +\alpha_1 \alpha_3 p_2^2
         + \alpha_2 \alpha_3 p_1^2 \right)^{\Sigma \nu_{i} - n/2}} .
\ee
If the sum of the powers of denominators,
$\sum \nu_i \equiv \nu_1+\nu_2+\nu_3$, is equal
to $n$, the triangle is ``unique'', and a very simple result can be
obtained from (\ref{Fp}) (see ref. \cite{VPH})
\be
\label{uniq1}
\left. \frac{}{}
J_3 (n; \; \nu_1,\nu_2,\nu_3) \right|_{\Sigma \nu_i = n}
=  \pi^{n/2} \; \mbox{i}^{1-n} \prod_{i=1}^{3}
\frac{\Gamma ( n/2 - \nu_i)}{\Gamma (\nu_i)} \;
\frac{1}{(p_i^2)^{n/2-\nu_i}} .
\ee
There are also two other useful relations of such type
(see ref.~\cite{UsKaz'83}):
\be
\label{uniq2}
\left. \frac{}{}  \!\!\!\!
J_3 (n; \; \nu_1,\nu_2,\nu_3) \right|_{\Sigma \nu_i = n+1}
=  \pi^{n/2} \; \mbox{i}^{1-n}
\sum_{j=1}^{3} \left\{ \prod_{i=1}^{3}
\frac{\Gamma ( n/2 - \nu_i -\delta_{ij}+ 1)}{\Gamma (\nu_i)} \;
\frac{1}{(p_i^2)^{n/2-\nu_i-\delta_{ij}+ 1}} \right\} ,
\ee
\[
\left\{ \left.   \frac{}{}
\nu_1 J_3 (n; \; \nu_1 +1,\nu_2,\nu_3)
+ \nu_2 J_3 (n; \; \nu_1 ,\nu_2 +1,\nu_3)
+ \nu_3 J_3 (n; \; \nu_1 ,\nu_2,\nu_3 +1) \right\}
\right|_{\Sigma \nu_i = n-2}
\]
%\nonumber \\
\be
\label{uniq3}
=  \pi^{n/2} \; \mbox{i}^{1-n} \prod_{i=1}^{3}
\frac{\Gamma ( n/2 - \nu_i -1)}{\Gamma (\nu_i)} \;
\frac{1}{(p_i^2)^{n/2-\nu_i-1}}.
\hspace{1cm}
\ee

For $\nu_1=\nu_2=\nu_3=1, \; n=4$ we define
\be
\label{defC1}
C^{(1)}(p_1^2, p_2^2, p_3^2) \equiv  J(4; 1, 1, 1)
= \frac{\mbox{i} \pi^2}{p_3^2} \; \Phi^{(1)}(x,y),
\ee
where the dimensionless variables $x$ and $y$ are
\be
\label{xy}
 x \equiv \frac{p_1^2}{p_3^2} \hspace{0.5cm}\mbox{   ,   }
\hspace{0.5cm} y \equiv \frac{p_2^2}{p_3^2} \; .
\ee
The function $\Phi^{(1)}(x,y)$ can be represented as a Mellin--Barnes
integral (see in \cite{Us'75}),
\be
\label{MBC1}
\Phi^{(1)}(x,y) =
\frac{1}{(2\pi \mbox{i})^2}
\Int_{-\mbox{i}\infty}^{\mbox{i}\infty}
\Int_{-\mbox{i}\infty}^{\mbox{i}\infty}
\mbox{d}u\; \mbox{d}v\; x^u\; y^v \;
\Gamma^2 (-u) \; \Gamma^2 (-v) \; \Gamma^2 (1+u+v) ,
\ee
or as the parametric integral,
\be
\label{Phi1int}
\Phi^{(1)} (x,y) = - \Int_0^1
\frac{\mbox{d} \xi}{y \xi^2 + (1-x-y) \xi +x} \;
\left( \ln\frac{y}{x} + 2\ln{\xi}  \right),
\ee
where the denominator can be also represented as a propagator,
\be
\label{den}
p_3^2 \; (y \xi^2 + (1-x-y) \xi + x) = (p_1 + \xi p_2)^2 .
\ee
The result in terms of dilogarithms is:
\be
\label{Phi1}
\Phi^{(1)} (x,y) = \frac{1}{\lambda} \left\{ \frac{}{}
2 \left( \Li{2}{-\rho x} + \Li{2}{-\rho y} \right)
+ \ln\frac{y}{x} \ln{\frac{1+\rho y}{1+\rho x}}
+ \ln(\rho x) \ln(\rho y) + \frac{\pi^2}{3}
\right\} ,
\ee
where
\be
\label{lambda}
\lambda(x,y) \equiv \sqrt{(1-x-y)^2 - 4 x y} \; \; \; ,
\; \; \; \rho(x,y) \equiv 2 \; (1-x-y+\lambda)^{-1} .
\ee
The results of the type of (\ref{Phi1}) are well-known (see, e.g.,
in ref.~\cite{tHV'79}). Here we use the notation of the papers
\cite{JPA,UD1,UD2}. It should be noted that the same function
$\Phi^{(1)}$ also occurs in the results for some other diagrams
(for example, the one-loop box diagram \cite{UD1,UD2} and two-loop
vacuum diagram with masses \cite{DT,DST}). Moreover, in four dimensions
one-loop diagrams with larger number of external lines can be
reduced to the four-point function (see, e.g., in refs.~\cite{pentagon}),
and therefore to a combination of $\Phi^{(1)}$ functions.

\vspace{3mm}

{\bf 3.} In this section we will present the results for the diagrams
shown in fig.~1(a,c). The ``ladder'' diagram (fig.~1a),
\be
\label{defC2}
C^{(2)}(p_1^2, p_2^2, p_3^2) = \int
\frac{\mbox{d}^4 r}{r^2 (p_1+r)^2 (p_2-r)^2}
C^{(1)}((p_1+r)^2, (p_2-r)^2, p_3^2)
\equiv \left( \frac{\mbox{i} \pi^2}{p_3^2} \right)^2
\; \Phi^{(2)}(x,y) ,
\ee
was calculated in \cite{UD1}. In this paper we will show that the result
for the diagram in fig.~1c can be expressed in terms of the same function.

Let us introduce analytic regularization of these diagrams by shifting
the powers of propagators by $\delta_i$ (as shown in fig.~2),
%===================================================================
%\input ud-fig2.tex
\begin{figure}
\setlength {\unitlength}{1mm}
\begin{picture}(150,70)(0,0)
\put (0,35) {\line(1,0){42}}
\put (18,35) {\line(1,1){30}}
\put (18,35) {\line(1,-1){30}}
\put (42,11) {\line(0,1){48}}
\put (0,31) {\makebox(0,0)[bl]{\large $p_3$}}
\put (50,64) {\makebox(0,0)[bl]{\large $p_1$}}
\put (50,3) {\makebox(0,0)[bl]{\large $p_2$}}
\put (17,50) {\makebox(0,0)[bl]{\large $1-\delta_1$}}
\put (43,46) {\makebox(0,0)[bl]{\large $1-\delta_2$}}
\put (17,15) {\makebox(0,0)[bl]{\large $1-\delta_2$}}
\put (43,20) {\makebox(0,0)[bl]{\large $1-\delta_1$}}
\put (4,37) {\makebox(0,0)[bl]{\large $1-\delta_3$}}
\put (26,37) {\makebox(0,0)[bl]{\large $1-\delta_3$}}
\put (3,35) {\circle*{1}}
\put (18,35) {\circle*{1}}
\put (42,35) {\circle*{1}}
\put (42,59) {\circle*{1}}
\put (42,11) {\circle*{1}}
\put (55,33) {\makebox(0,0)[bl]{\Large $= \; \prod\limits_{i=1}^{3} \;
                                        \frac{\Gamma^2 (1+\delta_i)}
                                             {\Gamma^2 (1-\delta_i)}$}}
\put (95,35) {\line(1,0){11}}
\put (106,35) {\line(2,1){50}}
\put (106,35) {\line(2,-1){50}}
\put (126,25) {\line(0,1){20}}
\put (146,15) {\line(0,1){40}}
\put (97,31) {\makebox(0,0)[bl]{\large $p_3$}}
\put (157,60) {\makebox(0,0)[bl]{\large $p_1$}}
\put (157,8) {\makebox(0,0)[bl]{\large $p_2$}}
\put (108,47) {\makebox(0,0)[bl]{\large $1+\delta_1$}}
\put (108,24) {\makebox(0,0)[bl]{\large $1+\delta_2$}}
\put (128,55) {\makebox(0,0)[bl]{\large $1+\delta_2$}}
\put (128,13) {\makebox(0,0)[bl]{\large $1+\delta_1$}}
\put (128,34) {\makebox(0,0)[bl]{\large $1+\delta_3$}}
\put (148,34) {\makebox(0,0)[bl]{\large $1+\delta_3$}}
\put (106,35) {\circle*{1}}
\put (126,25) {\circle*{1}}
\put (126,45) {\circle*{1}}
\put (146,15) {\circle*{1}}
\put (146,55) {\circle*{1}}
\end{picture}
\caption{}
\end{figure}
%======================================================================
provided
that $\delta_1+\delta_2+\delta_3=0$. Since these diagrams are convergent,
the limit $\delta_i \to 0$ should give results that do not depend
on the way how these $\delta$'s vanish.
By use of the ``uniqueness'' relations (\ref{uniq1}) and (\ref{uniq3}),
the diagram on the r.h.s. of fig.~2 can be reduced to one-loop integrals
(\ref{defJ}) (see in \cite{UD1}), and the result is
\bea
\label{uniqC2}
\frac{i \pi^2}{(p_3^2)^{1-\delta_3}}
\prod \frac{\Gamma(1- \delta_i)}{\Gamma(1+ \delta_i)}
\left\{ \frac{1}{\delta_1 \delta_2} J(4; \; 1, 1, 1+\delta_3) \right.
\hspace{5cm} \nonumber \\
\left.
+\frac{1}{\delta_1 \delta_3} (p_1^2)^{\delta_1} J(4; \; 1, 1, 1-\delta_2)
+\frac{1}{\delta_2 \delta_3} (p_2^2)^{\delta_2} J(4; \; 1, 1, 1-\delta_1)
\right\} .
\eea
Then, using the parametric representation
\be
\label{intPhi}
J(4; 1,1,1+\delta) = \frac{\mbox{i} \pi^2}{(p_3^2)^{1+\delta}} \;
\frac{1}{\delta} \Int_0^1 \mbox{d} \xi
\frac{(y \xi)^{-\delta} - (x/ \xi)^{-\delta}}
     {y \xi^2 + (1-x-y) \xi + x}
\ee
and considering the limit $\delta_i \to 0 \; (\sum \delta_i = 0)$, we
obtain \cite{UD1}
\be
\label{Phi2int}
\Phi^{(2)} (x,y) = - \frac{1}{2} \Int_0^1
\frac{\mbox{d} \xi}{y \xi^2 + (1-x-y) \xi +x} \;
\ln{\xi} \left( \ln\frac{y}{x} + \ln{\xi}  \right)
\left( \ln\frac{y}{x} + 2\ln{\xi}  \right) .
\ee
This integral can be evaluated as
\bea
\label{Phi2}
\Phi^{(2)} (x,y) =  \frac{1}{\lambda}
\left\{6 \left( \Li{4}{-\rho x} + \Li{4}{-\rho y} \right)
    + 3 \ln\frac{y}{x}  \left( \Li{3}{-\rho x} - \Li{3}{-\rho y} \right)
\right. \hspace{8mm}
\nonumber \\ \hspace{3cm}
 + \frac{1}{2} \ln^2 \frac{y}{x}
        \left( \Li{2}{-\rho x} + \Li{2}{-\rho y} \right)
 + \frac{1}{4} \ln^2 (\rho x) \ln^2 (\rho y)  \hspace{2cm}
\nonumber \\ \hspace{3cm}  \left.
 + \frac{\pi^2}{2} \ln (\rho x) \ln (\rho y)
 + \frac{\pi^2}{12} \ln^2 \frac{y}{x} +\frac{7 \pi^4}{60} \right\},
\hspace{3cm}
\eea
where the polylogarithms $\Li{N}{z}$ are defined as (see ref.~\cite{Lewin})
\be
\label{LiN}
\Li{N}{z} = \frac{(-1)^N}{(N-1)!} \Int_0^1  \mbox{d}\xi \;
\frac{\ln^{N-1} \xi}{\xi -z^{-1}} .
\ee
Note that in ref.~\cite{UD2,Bro-lad} these results (\ref{Phi2int}) and
(\ref{Phi2})
were generalized to the case of ladder diagrams with an arbitrary number of
loops. Moreover, it was shown that the four-point ladder diagrams can
be expressed in terms of the same functions as three-point ones
(for example, the result for the ``double box'' diagram contains the
same function $\Phi^{(2)}$).

Let us consider now the diagram on the l.h.s. of fig.~2 (it corresponds to
the diagram in fig.~1c). Using the relation (\ref{uniq2}) for the lines
that are connected with the ``internal'' vertex of this diagram, and
applying also the formula (\ref{uniq1}), it is easy to obtain the result
of the same form as (\ref{uniqC2}), multiplied by the factor
that is indicated in front of the r.h.s. of the equation in fig.~2.
Considering the limit $\delta_i \to 0 \; (\sum \delta_i =0)$ we find
that the result for the diagram in fig.~1c (at $n=4$) is
\be
\label{diag-c}
p_3^2 \; C^{(2)} (p_1^2, p_2^2, p_3^2) =
\frac{(\mbox{i} \pi^2)^2}{p_3^2} \; \Phi^{(2)}(x,y),
\ee
where $\Phi^{(2)}$ is defined by (\ref{Phi2int}), (\ref{Phi2}).
The equation  (\ref{diag-c}) (see fig.~2) can be easily checked by joining
external lines with the momenta $p_1$ and $p_2$ together and integrating
over the obtained loop momentum. The fact that both three-loop
propagator-type diagrams should give the result proportional to
$20 \zeta(5)$ is well-known \cite{BroUs} (the result for the
ladder graph was calculated in \cite{CKT,BU'83}).
It should be noted that the equation shown in fig.~2 can be generalized
to the case of larger number of loops.

\vspace{3mm}

{\bf 4.} In this section we shall examine the non-planar graph shown
in fig.~1b. The corresponding Feynman integral is defined as
\be
\label{defCc}
\widetilde{C}^{(2)}(p_1^2, p_2^2, p_3^2)
= \int\int \frac{\mbox{d}^n q \; \mbox{d}^n r}
                {q^2 \; r^2 \; (p_1-q)^2 \; (p_2+r)^2
                 \; (p_1-q-r)^2 \; (p_2+q+r)^2}
\ee
This function is symmetric with respect to all external lines, and each
single loop of this diagram corresponds to a four-point function.

To evaluate the crossed diagram $\widetilde{C}^{(2)}$, it is convenient
to use Fourier transform to the coordinate space. It is easy to show
that the ``topology'' of the diagram remains the same in the
$x$-space (see fig.~3):
%====================================================================
%\input ud-fig3.tex
\begin{figure}[bth]
\centering
\setlength {\unitlength}{0.7mm}
\begin{picture}(150,58)(0,0)
%\put (30,30) {\line(1,0){20}}
\put (50,30) {\line(3,1){60}}
\put (50,30) {\line(3,-1){60}}
\put (80,20) {\line(1,1){30}}
\put (80,40) {\line(1,-1){30}}
\put (50,30) {\circle*{2}}
\put (80,20) {\circle*{2}}
\put (80,40) {\circle*{2}}
\put (110,50) {\circle*{2}}
\put (110,10) {\circle*{2}}
%\put (22,30) {\makebox(0,0)[bl]{\large $p_3$}}
%\put (134,52) {\makebox(0,0)[bl]{\large $p_1$}}
%\put (134,7) {\makebox(0,0)[bl]{\large $p_2$}}
\put (47,34) {\makebox(0,0)[bl]{\large $x_3$}}
\put (107,54) {\makebox(0,0)[bl]{\large $x_1$}}
\put (107,1) {\makebox(0,0)[bl]{\large $x_2$}}
\put (77,44) {\makebox(0,0)[bl]{\large $z_1$}}
\put (77,11) {\makebox(0,0)[bl]{\large $z_2$}}
\end{picture}
\caption{}

\end{figure}
%=======================================================================
there are three points $x_1$, $x_2$,
$x_3$ (associated with external vertices), and there are
two internal points $z_1$ and $z_2$, each of them being connected with
all the three $x_i$. Note that in the $x$-space both integrations (with
respect to the positions of $z_1$ and $z_2$) are independent.
Therefore, the crossed diagram in fig.~3 factorizes (in the $x$-space),
and the Fourier transform of (\ref{defCc}) gives
\be
\label{Fourier}
- \frac{\mbox{i}}{(2\pi)^2}
\left\{ \int \frac{\mbox{d}^4 z}{z^2 \; (x_1-x_3-z)^2 \; (x_2-x_3-z)^2}
\right\}^2 .
\ee
Note that the integral in braces has the same structure as the
momentum-space one-loop integral (\ref{defJ}). So, we may apply the
Mellin--Barnes representation (\ref{MBC1}) to each of the two factorized
integrals in (\ref{Fourier}). After we have done it, we get
the following $x$-dependent structure in the integrand:
\be
\label{xxx}
\left[ (x_2-x_3)^2 \right]^{u_1+u_2}
\left[ (x_1-x_3)^2 \right]^{v_1+v_2}
\left[ (x_1-x_2)^2 \right]^{2-u_1-u_2-v_1-v_2} ,
\ee
where $u_1, v_1$ and $u_2,v_2$ are the Mellin--Barnes contour integral
variables. In the momentum space, this product (\ref{xxx}) corresponds
to the ``unique'' triangle
\[
J(4; 2-u_1-u_2, 2-v_1-v_2, u_1+u_2+v_1+v_2)
\]
that can be written as a product of the powers of $p_i^2$ (see eq.
(\ref{uniq1})). Finally, collecting all gamma functions occurring in the
transformations involved, we see that the resulting four-fold
Mellin--Barnes integral (in the momentum space) can be split
into two two-fold integrals, each of them corresponding to
the one-loop function (\ref{MBC1})! So, we obtain a very simple result:
\be
\label{Cc2C1}
\widetilde{C}^{(2)}(p_1^2, p_2^2, p_3^2)
= \left( C^{(1)}(p_1^2, p_2^2, p_3^2) \right) ^2
= \left( \frac{i \pi^2}{p_3^2} \right) ^2
  \left( \Phi^{(1)} (x,y) \right) ^2
\ee
with $\Phi^{(1)} (x,y)$ defined by (\ref{Phi1int})-(\ref{Phi1})
(therefore, $\widetilde{C}^{(2)}$ can be expressed
in terms of dilogarithms and their products).
We have mentioned this factorization property in ref.~\cite{YadFiz}.
This result (\ref{Cc2C1}) has also been checked by
joining two external lines and integrating over the loop,
that gives the correct result proportional to $20\zeta(5)$
(see ref.~\cite{CKT}).

\vspace{3mm}

{\bf 5.} The diagrams shown in fig.~1(d,e,f) are divergent, and we
shall employ dimensional regularization \cite{dimreg} to calculate
the singular and finite (in $\ep=(4-n)/2$) parts. All these diagrams
contain one-loop propagator ``bubbles'' that can be easily
calculated as
\be
\label{bubble}
\frac{\mbox{i}^{1+2\ep} \pi^{2-\ep}}{(p^2)^{\ep}} \;
\frac{\Gamma^2(1-\ep) \; \Gamma(\ep)}{\Gamma(2-2\ep)}
= \frac{\mbox{i}^{1+2\ep} \pi^{2-\ep}}{(p^2)^{\ep}} \;
\Gamma(1+\ep) \left[ \frac{1}{\ep} + 2 - \ep \frac{\pi^2}{6}
+ {\cal O}(\ep^2)  \right] ,
\ee
where $p$ is the ``external'' momentum corresponding to the ``bubble'',
and we put the dimensional regularization scale parameter $\mu_0 = 1$.
The resulting integrals correspond to eq.~(\ref{defJ}),
and they are: $J(4-2\ep;1,1,1)$ (fig.~1d), $J(4-2\ep;1,1,1+\ep)$ (fig.~1e)
and $J(4-2\ep;1,1,\ep)$ (fig.~1f). The problem is that these
triangle integrals should be evaluated up to their $\ep$-parts
(because they are multiplied by (\ref{bubble}) containing $1/\ep$).

We start with the diagram shown in fig.~1d. We employ
the ``uniqueness'' relation (\ref{uniq1}) to make the following
transformations:
\bea
\label{transf}
J(4-2\ep;1,1,1)
= \frac{\Gamma (1+\ep) \Gamma^3 (1-\ep)}{\Gamma(1-2\ep)} \;
                  (p_3^2)^{-\ep} \;
                  J(4-2\ep;1-\ep,1-\ep,1+\ep)
\nonumber \\
= (p_1^2)^{-\ep} \; (p_2^2)^{-\ep} \;
                  J(4-2\ep;1,1,1-2\ep) .
\hspace{2cm}
\eea
Using these formulae and the Feynman
parameters (\ref{Fp}), we can obtain the following one-parametric
integral representation:
\be
\label{int-dr}
J(4-2\ep;1,1,1) = \frac{\pi^{2-\ep}\;\mbox{i}^{1+2\ep}}{(p_3^2)^{1+\ep}} \;
                 \frac{\Gamma (1+\ep) \Gamma^2 (1-\ep)}{\Gamma(1-2\ep)} \;
                 \frac{1}{\ep} \Int_0^1
 \frac{\mbox{d}\xi\;\xi^{-\ep} \left( (y\xi)^{-\ep} - (x/\xi)^{-\ep} \right)}
      {\left( y \xi^2 + (1-x-y)\xi + x \right)^{1-\ep}} .
\ee
{}From here, we immediately get a correct result (\ref{defC1}), (\ref{Phi1int})
for the case of $\ep=0$.
Note that the denominator in (\ref{int-dr}) is the same as in
(\ref{Phi1int}) and (\ref{intPhi}), and it can be represented as
(\ref{den}) (this makes it possible to use (\ref{int-dr}) as a ``block''
in multiloop calculations).

Let us denote
\be
\label{J111}
J(4-2\ep;1,1,1) = \frac{\pi^{2-\ep}\;\mbox{i}^{1+2\ep}}{(p_3^2)^{1+\ep}} \;
                 \Gamma (1+\ep) \;
    \left\{ \Phi^{(1)}(x,y) + \ep \; \Psi^{(1)}(x,y) + {\cal O}(\ep^2)
\right\},
\ee
where $\Phi^{(1)}$ is defined by (\ref{Phi1int}), (\ref{Phi1}).
The parametric representation for $\Psi^{(1)}$ can be obtained
from (\ref{int-dr}), and it is
\bea
\label{Psiint}
\Psi^{(1)}(x,y) = - \Int_0^1 \frac{\mbox{d} \xi}
{y \xi^2 + (1 \! - \! x \! - \! y) \xi + x} \;
\left\{ \left( \ln\frac{y}{x} + 2\ln\xi \right)
        \ln(y \xi^2 + (1 \! - \! x \! - \! y) \xi + x) \right.
\nonumber \\
\left. -2 \ln y \ln\xi - 2 \ln^2\xi - \frac{1}{2} \ln(xy) \ln\frac{y}{x}
\right\} .  \hspace{1cm}
\eea
This integral can be evaluated in terms of polylogarithms (\ref{LiN})
(up to the third order), and we get
\bea
\label{Psi1}
\Psi^{(1)}(x,y) = - \frac{1}{\lambda}
\left\{  4 \; \Li{3}{-\frac{\rho x (1+ \rho y)}{1- \rho^2 x y}}
       + 4 \; \Li{3}{-\frac{\rho y (1+ \rho x)}{1- \rho^2 x y}}
       - 4 \; \Li{3}{\frac{ -\rho^2 x y}{1- \rho^2 x y}}
\right. \nonumber \\
       + 2 \; \Li{3}{\frac{\rho x (1+ \rho y)}{1+ \rho x}}
       + 2 \; \Li{3}{\frac{\rho y (1+ \rho x)}{1+ \rho y}}
       - 2 \; \Li{3}{\rho^2 x y} - 2 \zeta(3)
\nonumber \\
       - 2 \ln y \; \Li{2}{\frac{\rho x (1+ \rho y)}{1+ \rho x}}
       - 2 \ln x \; \Li{2}{\frac{\rho y (1+ \rho x)}{1+ \rho y}}
\hspace{24mm}
%       + \frac{\pi^2}{3} \ln(\rho^2 x y)
\nonumber \\
       -\frac{2}{3} \ln^3 (1- \rho^2 x y)
       +\frac{2}{3} \ln^3 (1+ \rho x) + \frac{2}{3} \ln^3 (1+ \rho y)
       + 2 \ln\rho \ln^2 (1- \rho^2 x y)
\nonumber \\
       - 2 \ln(1\! - \!\rho^2 x y)
         \left( \ln(\rho x) \ln(\rho y)
                + \ln\frac{y}{x} \ln\frac{1\!+\!\rho y}{1\!+ \!\rho x}
                + 2 \ln(1 \!+\! \rho x) \ln(1\!+\! \rho y)
                + \frac{\pi^2}{3} \right)
\nonumber \\  \left.
       +\frac{1}{2} \ln(\rho^2 x y)
        \left( \ln(\rho x) \ln(\rho y)
              + \ln\frac{y}{x} \ln \frac{1+ \rho y}{1+ \rho x}
              - \ln^2 \frac{1+ \rho y}{1+ \rho x}
         + \frac{2 \pi^2}{3} \right) \right\} ,
\eea
where $\lambda$ and $\rho$ are defined by (\ref{lambda}).
This expression is explicitly symmetric in $x$ and $y$.
Note that in ref.~\cite{NMB} the massive case of such diagram was
considered, and the result for the $\ep$-part also contained
trilogarithms. By use of the integration-by-parts technique \cite{ibp},
we can obtain the results for the integrals $J(4-2\ep; N_1, N_2, N_3)$
(with integer $N_1, N_2, N_3$) in terms of the same functions
\cite{JPA} (moreover, (\ref{J111}) is the only integral of such type where
$\Psi^{(1)}$ occurs in the $\ep$-part, other integrals will contain
$\Phi^{(1)}$ only).

Another integral that we need to know up to the $\ep$-part, is
$J(4-2\ep; 1,1,1+\ep)$ (it corresponds to the diagram in fig.~1e).
If we write Feynman parametric representation
for this integral, we may observe, that the $\ep$-part of the
parametric integral can be represented as a linear combination
of $\ep$-parts of the integrals $J(4-2\ep; 1,1,1)$ and
$J(4; 1,1,1+\ep)$. The result for the latter integral follows
from eq.(\ref{intPhi}) and was obtained in \cite{UD1} (it can be
expressed via the $\Phi^{(1)}$-function). Finally, we get
\be
\label{int(1+ep)}
J(4-2\ep; 1,1,1+\ep) =
\frac{\pi^{2-\ep}\;\mbox{i}^{1+2\ep}}{(p_3^2)^{1+2\ep}}
                 \Gamma (1+\ep)
    \left\{ \Phi^{(1)}(x,y) \left( 1 - \frac{\ep}{2} \ln(xy) \right)
           + \ep \Psi^{(1)}(x,y) + {\cal O}(\ep^2) \right\},
\ee
with the same $\Psi^{(1)}$ as in (\ref{J111}).

The result for $J(4-2\ep; 1,1,\ep)$ (corresponding to the
diagram in fig.~1f) can be obtained either by direct
evaluation (like (\ref{int-dr})) or in a way similar to the evaluation
of (\ref{int(1+ep)}). It is divergent, but does not contain the
$\Psi^{(1)}$ function:
\bea
\label{int(ep)}
J(4-2\ep; 1,1,\ep) =
\frac{\pi^{2-\ep}\;\mbox{i}^{1+2\ep}}{(p_3^2)^{2\ep}} \;
                 \Gamma (1+\ep) \; \frac{1}{2 (1-3\ep)}
\hspace{5cm}  \nonumber \\
\times \left\{ \frac{1}{\ep} - \ep \; \ln x \ln y
               + \ep \; (1-x-y) \; \Phi^{(1)}(x,y) - \ep \; \frac{\pi^2}{6}
               + {\cal O}(\ep^2) \right\} .
\eea
The results (\ref{int(1+ep)}) and (\ref{int(ep)}) enable one
to evaluate other integrals
$J(4-2\ep; N_1, N_2, N_3+\ep)$ (with integer $N_1, N_2, N_3$)
by using the formulae of the paper \cite{JPA}. For example,
\bea
\label{int(2+ep)}
J(4-2\ep; 1,1, 2+\ep) =
\frac{\pi^{2-\ep}\;\mbox{i}^{1+2\ep}}{(p_3^2)^{2+2\ep}} \;
                 \Gamma (1+\ep) \; \frac{1}{2 (1 +\ep)} \; \frac{1}{x \; y}
\hspace{4cm}  \nonumber \\
\times \left\{ - \frac{1}{\ep} + 2 \ln x + 2 \ln y
               - \ep \; (2 \ln^2 x + 2 \ln^2 y + \ln x \ln y) \right.
\hspace{1cm}  \nonumber \\
       \left.  - 3 \ep \; (1-x-y) \; \Phi^{(1)}(x,y) + \ep \; \frac{\pi^2}{6}
               + {\cal O}(\ep^2) \right\}.
\eea
It should be noted that the $1/\ep$ pole in (\ref{int(ep)}) has
an ultraviolet origin, while the $1/\ep$ pole in (\ref{int(2+ep)}) has
an infrared origin.

\vspace{3mm}

{\bf 6.} In this paper, we have presented exact results for all the types
of three-point off-shell contributions shown in fig.~1(a--f).
An interesting fact is that some diagrams can be expressed in terms
of the functions corresponding to other diagrams. For example, the
results for the diagrams shown in fig.~1(a,c) involve the same function
(see also fig.~2), and the result for the non-planar diagram in
fig.~1b can be represented as the one-loop triangle integral squared.
There are only three independent functions occurring in the results
(up to the two-loop level), namely: $\Phi^{(1)}$ (\ref{Phi1}) (in fact,
this function is connected with the one-loop triangle), $\Phi^{(2)}$
(\ref{Phi2}) and $\Psi^{(1)}$ (\ref{Psi1}). All these functions
are evaluated in terms of polylogarithms (\ref{LiN}).

The presented results are necessary for constructing asymptotic expansions
of two-loop three-point massive diagrams in the limit of large
external momenta (see, e.g., in \cite{Smi-book} and references therein).
The two-loop two-point case of such expansion has been considered
in ref.~\cite{DST}.
They can also be used for evaluating some radiative corrections
(like two-loop off-shell contribution to the three-gluon vertex).
We should remember, however, that some scalar products (that may
occur in the numerator) cannot be represented in terms of the
denominators involved, and such terms should be considered separately.
It should be noted that the on-shell singularities (that occur when
some of the external momenta vanish) can be parametrized by putting
$p_i^2 = \mu^2$ (for those momenta that vanish), and they will appear
as powers of $\ln\mu$ (up to the fourth order). If dimensional
regularization is used to regulate these on-shell singularities,
they appear as poles in $\ep$ up to $1/\ep^4$  \cite{on-shell}.

\vspace{2mm}

{\bf Acknowledgements.} N.U. is grateful, for hospitality, to
the Department of Physics, University of Bergen where the final
part of this work was done.
We are grateful to P.~Osland for his interest in this work,
and for creating favourable conditions needed for completing it.
The results presented in this paper
are also connected with the continuation of a project that has been
started in Leiden, mainly due to F.A.~Berends. We are thankful to
D.J.~Broadhurst and U.~Nierste for useful comments.
The research was supported by the Research Council of Norway.

\end{document}